\documentclass[showpacs,preprintnumbers,amsmath,amssymb,prb,superscriptaddress]{revtex4}
\usepackage{graphicx}
\usepackage{dcolumn}
\usepackage{bm}

\begin{document}

\newcommand {\be}{\begin{equation}}
\newcommand {\ee}{\end{equation}}
\newcommand {\bea}{\begin{eqnarray}}
\newcommand {\eea}{\end{eqnarray}}
\newcommand {\nn}{\nonumber}

\title{Anatomy of Gossamer Superconductivity} 

\author{Stephan Haas}   
\affiliation{Department of Physics and Astronomy, University of Southern
California, Los Angeles, CA 90089-0484}

\author{Kazumi Maki}  
\affiliation{Department of Physics and Astronomy, University of Southern
California, Los Angeles, CA 90089-0484}

\author{Thomas Dahm}  
\affiliation{Institut f\"ur Theoretische Physik, Universit\"at T\"ubingen,
Auf der Morgenstelle 14, D-72706 T\"ubingen, Germany}

\author{Peter Thalmeier} 
\affiliation{Max-Planck Institute for Chemical Physics of Solids, N\"othnitzer
Str. 40, D-01187 Dresden, Germany}

\date{\today}

\pacs{PACS numbers: }

\begin{abstract}
There are many systems in which two order parameters compete with each 
other. Of particular interest are systems in which these order parameters
are both unconventional. In this contribution, we examine the representative
example of a d-wave superconductor in the presence of a d-wave density wave,
which has been suggested as a model for the pseudogap phase in the high-$\rm
T_c$ superconductors. The physical properties of unconventional 
superconductivity in the presence of an anisotropic charge density wave are
investigated within mean field theory. This model describes many features 
that were anticipated by an earlier phenomenological treatment of Tallon
and Loram. In addition, the quasiparticle density of states in the presence
of these two order parameters is calculated, which should be accessible by 
scanning tunneling microscopy.

\end{abstract}

\pacs{} 
 
\maketitle

\section{Introduction}

The discovery of unconventional high-temperature superconductivity by
Bednorz and M\"uller\cite{bednorz} in 1986 took many in the community 
by surprise. Subsequent developments and some of the initial confusions 
are documented in the textbook by Enz.\cite{enz} Among many others, 
Anderson\cite{anderson} pointed out in 1987 that high-$\rm T_c$ 
superconductivity should be understood in terms of a competition between a
Mott insulating state in the limit of zero hole doping and a superconducting
state at optimal doping. Meanwhile, experimental evidence for
the d-wave nature of high-$\rm T_c$
superconductivity was found,\cite{damascelli,harlingen,tsuei}
and general features of the phase diagram were established.\cite{tallon}
However, the nature of the pseudogap phase is still being debated. 
However, now there is increasing consensus that this regime can be described by
in terms of a d-density-wave.\cite{benfatto,chakravarty,dora}

Recently, Laughlin
proposed a very intuitive formulation regarding
how to deal with superconductors
close to a Mott insulating state.
\cite{laughlin}. This ``gossamer
superconductivity" is characterized by a very small superfluid density and
spectral weight of the superconducting quasiparticles. \cite{zhang}
In this contribution, we will
apply Laughlin's idea to a more general situation:
unconventional superconductivity in the presence of unconventional
density wave (UDW) order. More precisely, for high-$\rm T_c$ cuprates we
study d-wave
superconductivity in the presence of a d-density wave. \cite{dora}
This could either be a charge density wave or a spin density wave 
order parameter, although the distinction between these two possibilities
is not of importance for the following discussion. 
Here, we wish to examine whether some
features of the pseudogap phase can follow from the proposed model,
as it was suggested
by Tallon and Loram\cite{tallon} based on a phenomenological approach.
Furthermore, the recently observed weakening of superconductivity 
in the $\rm \kappa - (ET)_2$ salts when the sample is rapidly cooled
may be interpretetd within the model we are presenting in this paper. 
\cite{pinteric}

When both the superconductivity and the density wave are conventional,
this competition is not of great interest as long as the system can be reduced
to a single band or a single Fermi surface. In this case, there is typically
little room for coexistence,
and one of the order parameters, usually the superconductivity, dominates
over the
the other one.\cite{yamaji} A similar statement holds if only one
of the two competing order parameters is conventional.
However, the situation changes completely
when two unconventional order parameters  are considered \cite{thalmeier}.
If we look at heavy fermion systems and organic superconductors, we
see signs for
gossamer superconductivity almost everywhere, for example
in CeCu$_2$Si$_2$ \cite{thalmeier}, URu$_2$Si$_2$
\cite{ikeda,virosztek}, UPd$_2$Al$_3$, CeRhIn$_5$, CeCoIn$_5$
\cite{izawa,bianchi},
and the $\kappa$-(ET)$_2$ salts \cite{pinteric,miyagawa}. Therefore gossamer
superconductivity
as defined above may be a pervading feature in high-$\rm T_c$ cuprates,
heavy fermion superconductors and organic superconductors.

Here, we report calculations for nodal superconductivity in the
presence of nodal density wave order. For simplicity, we limit ourselves
to quasi-two-dimensional systems with cylindrical Fermi surfaces. 
We denote the superconducting and density wave order parameters by
$\Delta_1(\vec{k})$ and $\Delta_2(\vec{k})$, respectively, with 
angular dependence $\Delta_1(\vec{k}) = \Delta_1 f(\vec{k})$
and $\Delta_2(\vec{k}) = \Delta_2 g(\vec{k})$, where $\Delta_1$
and $\Delta_2$ are the maximum values of the corresponding
order parameters. Further, we limit ourselves to $f=\cos (2 \phi)$
(i.e. $d_{x^2-y^2}$-wave) while for $g$ we consider the two cases
(a) $g=\cos (2 \phi)$ and (b) $g=\sin (2 \phi)$. The first case
describes many features of the superconductivity in the pseudogap
phase in  high-$T_c$ cuprates, while we conjecture that the second
case applies for the glassy phase of $\kappa$-(ET)$_2$X with
X=Cu[N(CN)$_2$]Br.

\section{Quasiparticle Green Function}

Let us consider a one-band quasi-two-dimensional Fermi surface
where a nodal superconductor and a nodal density wave coexist. As a model
Hamiltonian we can think of a generalized Hubbard model \cite{dora2}.
The Green function is given in $4 \times 4$ spinor space as
\begin{equation}
G^{-1} ( \vec{k}, E ) = E - \xi \rho_3 \sigma_3 - \eta \rho_3
- \Delta_1 f \sigma_1 \rho_3 - \Delta_2 g \rho_1 \sigma_3
\end{equation}
where the Pauli matrices $\rho_i$ and $\sigma_i$ are operating on the
particle-hole space and the $\vec{k}$ and $\vec{k} \pm \vec{Q}$ space,
respectively, where $\vec{Q}$ is the nesting vector for the nodal
density wave. We have assumed here that the UDW is a charge density
wave. The
Green function possesses poles at the quasiparticle energies
\begin{equation}
E = \pm \sqrt{\left( \sqrt{\xi^2 + \Delta_2^2 g^2} 
\pm \eta \right)^2 + \Delta_1^2 f^2}
\end{equation}
which is well known in the literature. \cite{chakravarty,thalmeier} Here,
$\eta$ is the imperfect nesting term. 
Less well-known, perhaps, is the quasiparticle density of states
(DOS)  given by
\begin{widetext}
\begin{equation}
\frac{N \left( E \right)}{N_0} \equiv  G \left( E \right) = 
\frac{1}{2} \sum_\pm \mbox{Re} 
\left\{
\left\langle \frac{|E|}{\sqrt{\left( \sqrt{E^2-\Delta_1^2 f^2} \pm \eta
\right)^2
- \Delta_2^2 g^2}} 
 \right\rangle
\right\},
\end{equation}
\end{widetext}
where the sum has to be taken over the two terms with $\pm \eta$. Here,
$\langle .... \rangle$ denotes the angular average over $\phi $. 
In the limit $\eta \rightarrow 0$ this expression
 reduces to a very familiar form
\begin{equation}
G \left( E \right) = \mbox{Re } \left\langle 
\frac{\left| E \right|}{\sqrt{E^2-\Delta_1^2 f^2-\Delta_2^2 g^2}} \right\rangle
\end{equation}
In particular, for case (a) when $g=f=\cos(2\phi)$ we have
\begin{equation}
G \left( E \right) = \left\{ \begin{array}{cl} 
\frac{2}{\pi} x K \left( x \right) & \mbox{for } x < 1 \\ 
\frac{2}{\pi} K \left( x^{-1} \right) & \mbox{for } x > 1 
\end{array} \right. 
\end{equation}
where $x=E/\sqrt{\Delta_1^2 + \Delta_2^2}$ and $K$ is the complete elliptic
integral of the first kind. This has the same form as in $d$-wave
superconductors,\cite{won} with the only exception that the $d$-wave gap has to
be
replaced by the total gap $\sqrt{\Delta_1^2 + \Delta_2^2}$. On the other hand,
for case (b)
when $f=\cos(2\phi)$ and $g=\sin (2 \phi)$ the DOS develops a complete gap below
$E=\min(\Delta_1,\Delta_2)$ and we find
\begin{equation}
G \left( E \right) = \left\{ \begin{array}{cl} 
0 & \mbox{for } E < \Delta_1 \\
\frac{2}{\pi} \frac{\left| E \right|}{\sqrt{E^2-\Delta_1^2}} y K \left( y
\right) & \mbox{for } y < 1 \\ 
\frac{2}{\pi}  \frac{\left| E \right|}{\sqrt{E^2-\Delta_1^2}}
K \left( y^{-1} \right) & \mbox{for } y > 1 
\end{array} \right. 
\end{equation}
where $y=\sqrt{(E^2-\Delta_1^2)/(\Delta_2^2-\Delta_1^2)}$. Here we have assumed
$\Delta_1<\Delta_2$. The corresponding DOS 
with $\Delta_1/\Delta_2=$ 0.1, 0.3, and 0.5
is shown in Fig. 1(b). Except for the opening of the energy gap with increasing
$\Delta_1$, the DOS looks very similar to the case of $\Delta_1$ =0, with
a characteristic logarithmic singularity at $ |E| = \Delta_2$. 

\begin{figure}[h]
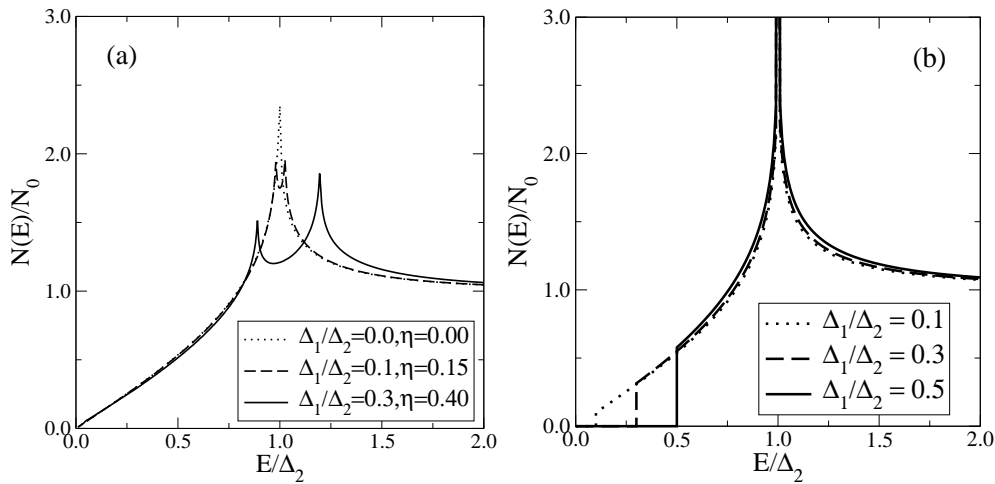

\includegraphics[width=6.5cm]{fig2}
\includegraphics[width=6.5cm]{fig1}
\caption{ Density of states of a gossamer superconductor, (a) with 
$f = g = \cos( 2 \phi )$, and (b) with $f = \cos (2 \phi )$ and 
$g = \sin (2 \phi )$.
 }
\end{figure}

As we shall see in the following, for case (a) a non-vanishing $\eta$ 
is crucial for the presence of two order parameters. In Fig. 1(a)
we show the corresponding DOS for the choices $\Delta_1$ = 0 ($\eta =0$),
$\Delta_1$ = 0.1$\Delta_2$ ($\eta = 0.15 \Delta_2$), and 
$\Delta_1$ = 0.3$\Delta_2$ ($\eta = 0.4 \Delta_2$). Due to the presence of
two order parameters, the logarithmic singularity at $ |E| = \Delta_2$
splits into two singularities, which are now located at 
$ |E| = \sqrt{(\Delta_2 \pm \eta )^2 + \Delta_1^2 }$. We expect this 
double-peak structure to be accessible to scanning tunneling microscope
measurements in the pseudogap phase of the high-$\rm T_c$ cuprates. 

Very recently, new measurements of the DOS in underdoped Bi2212 were
reported.\cite{yurgens} Indeed, they show a clear double-peak structure. 
Unfortunately, it was pointed out later that the secondary peak is 
a likely artifact of the heating process.\cite{zavaritsky,yurgens2}
However, based on the present model calculation we conclude that a 
second peak should indeed be visible in a careful experiment. 

\section{Gap Equations}

For case (a) with $ f = g$, the gap equations are given by
\begin{eqnarray}
\lambda_1^{-1} &=& 2 \pi T \sum_{n}  \mbox{Re} \Bigg\langle 
\frac{2 f^2}{\sqrt{\left( \sqrt{\omega_n^2 +
\Delta_1^2 f^2} + i \eta \right)^2 + \Delta_2^2 f^2}} 
\left( 1 + \frac{i \eta}{\sqrt{\omega_n^2 +\Delta_1^2 f^2}}
\right) \Bigg\rangle \\
\lambda_2^{-1} &=& 2 \pi T \sum_{n}  \mbox{Re} \Bigg\langle
\frac{2 f^2}{\sqrt{\left( \sqrt{\omega_n^2 +
\Delta_1^2 f^2} + i \eta \right)^2 + \Delta_2^2 f^2}}
\Bigg\rangle 
\label{gapeq}
\end{eqnarray}
for $\Delta_1$ and $\Delta_2$ respectively. Here, $\lambda_1$ and 
$\lambda_2$ are the dimensionless coupling constants for superconductivity
and density wave. Also, it is assumed here that $\lambda_1 < \lambda_2$, 
because otherwise there is no chance for the UDW to occur. Furthermore,
for the survival of a finite superconducting component we also need
$\eta \neq 0$. Then, for small $\eta$ one finds
$\Delta_1(0)/\Delta_2(0) \sim (\eta / \Delta_2(0))^{3/2}$, where 
$\Delta_1(0)$ and $\Delta_2(0)$ are the corresponding order parameters 
in the zero-temperature limit. 
Moreover, for the energy gap that can be observed by scanning tunneling
microscopy we find at low temperatures
$0.5 (\sqrt{(\Delta_2 + \eta)^2 + \Delta_1^2} +
\sqrt{(\Delta_2 - \eta)^2 + \Delta_1^2}) \simeq 
\Delta_2 ( 1 + \Delta_1^2/(2(\Delta_2^2 - \eta^2 )))$. Thus, in the limit
$\Delta_1 / \Delta_2 \ll 1$ and $|\eta |/\Delta_2 \ll 1$, scanning tunneling
microscopy will still identify $\Delta_2(0)$. 
Therefore, the puzzling result that the detected gap $\Delta (0) \simeq
2.14 \rm T_c$, where $\rm T_c$ is the transition temperature of the 
pseudogap phase, can be naturally interpreted in terms of a d-density
wave.\cite{dora2,oda,kugler} Finally, since the corresponding superfluid 
density at zero-temperature is given by
\begin{equation}
\rho_S (0,\eta ) = \frac{\Delta_1^2 (0)}{\Delta_2^2 (0) + \Delta_1^2 (0)},
\end{equation} 
there should be a substantial reduction of this quantity in the 
coexistence regime.

For the case (b) where $f \neq g$ the gap equations are now given by
\begin{eqnarray}
\lambda_1^{-1} &=& 2 \pi T \sum_{n}  \mbox{Re} \Bigg\langle 
\frac{2 f^2}{\sqrt{\left( \sqrt{\omega_n^2 +
\Delta_1^2 f^2} + i \eta \right)^2 + \Delta_2^2 g^2}} 
\left( 1 + \frac{i \eta}{\sqrt{\omega_n^2 +\Delta_1^2 f^2}}
\right) \Bigg\rangle \\
\lambda_2^{-1} &=& 2 \pi T \sum_{n}  \mbox{Re} \Bigg\langle
\frac{2 f^2}{\sqrt{\left( \sqrt{\omega_n^2 +
\Delta_1^2 f^2} + i \eta \right)^2 + \Delta_2^2 g^2}}
\Bigg\rangle 
\label{gapeq}
\end{eqnarray}
In contrast to case (a), here we always encounter coexistence of two 
order parameters, even when $\eta = 0$. In particular, for the case 
$\eta = 0$, Eqs. 10 and 11 can be solved at zero temperature.
\begin{eqnarray}
\Delta_2 (0) + \Delta_1 (0) & = & 2 \sqrt{E_1 E_2} \exp \left[ 
-(\lambda_1^{-1} + \lambda_2^{-1} )/2 \right] \\
\frac{\Delta_2 (0) - \Delta_1 (0)}{\Delta_2 (0) + \Delta_1 (0)} & = &
\lambda_1^{-1} - \lambda_2^{-1} ,
\end{eqnarray}
where $E_1$ and $E_2$ are cut-off energies.
Furthermore, the zero-temperature superfluid density is given by
\begin{eqnarray}
\rho_s (0,0) = \frac{\Delta_1 (0)}{\Delta_2 (0) + \Delta_1 (0) }.
\end{eqnarray}
Although the superfluid density is reduced in the present case,
the reduction is not as severe as it is for case (a).

\section{Concluding Remarks}

In this paper, we have analyzed the properties of a nodal superconductor
in the presence of a nodal density wave. Our preliminary results look
very consistent with experimental observations that have been reported for the 
pseudogap region of the high-$\rm T_c$ cuprates. For example, in contrast 
to the large quasiparticle energy gap observed by scanning tunneling 
microscopy, the corresponding superfluid density and the specific heat 
associated with superconductivity are much reduced.\cite{tallon}  The present
model calculation appears to describe these features consistently, although
more measurements on this issue are clearly necessary. The application of
this model to other potential ``gossamer superconductors" is promising.

\section{Acknowledgements} 

We have greatly benefited from discussions with Balazs Dora, Attila
Virosztek, Yuji Matsuda, Marko Pinteri\'c, Silvia Tomi\'c, and 
Hyekyung Won. TD and KM are grateful for the hospitality of the Max-Planck
Institute for Chemical Physics of Solids and the Max-Planck Institute for
the Physics of
Complex Systems at Dresden, where parts of this work were performed. 
SH acknowledges financial support by the National Science Foundation
DMR-0089882.

\end{document}